\documentclass{jaa}

\usepackage{graphicx,graphics,amsmath}
\usepackage{xcolor}
\usepackage[normalem]{ulem}

 \usepackage[sort&compress,round,comma,authoryear]{natbib}
\begin{document}

\title{AGN feedback with the Square Kilometer Array (SKA) and implications for cluster physics and cosmology}


\author{Asif Iqbal\textsuperscript{1,*}, Ruta Kale\textsuperscript{2}, Subhabrata Majumdar\textsuperscript{3}, Biman B. Nath\textsuperscript{4}, Mahadev Pandge\textsuperscript{5},  Prateek Sharma\textsuperscript{6}, Manzoor A. Malik\textsuperscript{1} \and
  Somak Raychaudhury\textsuperscript{7} }
\affilOne{\textsuperscript{1}Department of Physics, University of Kashmir, Hazratbal, Srinagar, J\&K, 190011, India\\}
\affilTwo{\textsuperscript{2}National Centre for Radio Astrophysics, Pune, India\\}
\affilThree{\textsuperscript{3}Tata Institute of Fundamental Research,  Mumbai, 400005, India\\}
\affilFour{\textsuperscript{4}Raman Research Institute, Bangalore, 560080, India\\}
\affilFive{\textsuperscript{5}Department of Physics, Dayanand Science College, Latur, 413512, India\\}
\affilSix{\textsuperscript{6}Indian Institute of Science, Bangalore, 560080, India\\}
\affilSeven{\textsuperscript{7}Inter-University Centre for Astronomy \& Astrophysics, Pune, India}


\twocolumn[{

\maketitle

\corres{asifiqbal@kashmiruniversity.net}


\begin{abstract}
AGN  feedback is regarded as an important non-gravitational process in galaxy clusters, providing useful 
constraints on large-scale structure formation. It modifies the structure and energetics of the intra-cluster medium (ICM) and hence its understanding is crucially needed in order to use clusters as high precision cosmological probes. In this context, particularly keeping in mind the upcoming high quality radio data expected from radio surveys like SKA with its higher sensitivity, high spatial and spectral resolutions, we review our current understanding of AGN feedback, its cosmological implications and the impact that SKA can have in revolutionizing our understanding of AGN feedback in large-scale structures. Recent developments regarding the AGN outbursts and its possible contribution to excess entropy in the hot atmospheres of groups and clusters, its correlation with the feedback energy in ICM, quenching of cooling flows and the possible connection between cool core clusters and radio mini-halos, are discussed. We describe current major issues regarding modeling of AGN feedback and its impact on the surrounding medium.  With regard to the future of AGN feedback studies, we  examine  the possible breakthroughs that can be expected from SKA observations. In the context of cluster cosmology, for example,  we point out the importance of SKA observations for cluster mass calibration by noting that  most of  $z>1$ clusters discovered by eROSITA X-ray mission can be expected to be followed up through a 1000 hour SKA-1 mid programme. Moreover, approximately $1000$ radio mini halos  and $\sim 2500$ radio halos at $z<0.6$ can be potentially detected by SKA1 and SKA2 and used as tracers of galaxy clusters and determination of cluster selection function.  
\end{abstract}

\keywords{Galaxy clusters---Cosmology---Intra-cluster medium---X-rays---Radio sources---Cooling flows.}

}]


\doinum{\#\#\#\#}
\artcitid{\#\#\#\#}
\volnum{\#\#\#\#}
\year{\#\#\#\#}

\section{Introduction}
Clusters of galaxies are the largest  virialized  objects in the universe and the youngest ones in the hierarchical scenario of structure formation.
This makes them ideal probes of the large-scale structure of the universe and, hence, of cosmological parameters. For example,
 the abundance of galaxy clusters provides sensitive constraints on the cosmological parameters that govern the growth of structures in the universe \citep{0,1,2,3,4,5,Bocquet2015}. In the scenario of hierarchical structure formation, galaxy clusters are formed from the 
merger of smaller units (galaxies, groups and small clusters) due to the gravitational collapse. About 90\% of the total baryons in galaxy clusters 
 are in the form of a hot ($10^7-10^8$ K) diffuse plasma, called the intra-cluster medium (ICM) which radiates in X-ray band via thermal bremsstrahlung. 
By observing its X-ray emission, one can study the thermodynamic properties of the ICM, which in turn can help us determine the dark matter profile in galaxy clusters.

High quality X-ray data have provided a number of interesting details regarding the thermodynamic properties of the ICM. For example, it has been found that there is a  
break in the self-similar scaling relations predicted by pure gravitational collapse of dark matter halos \citep{Kaiser1986,Edge1991,Andreon2011,Bonamente2008,Battaglia2012}. Similarly an excess entropy\footnote{Entropy  is defined as $K_g(r)=k_BTn_e(r)^{-2/3}$, where $n_e$ is the electron density and $k_B$ is the Boltzmann constant. It  is related to thermodynamic definition of specific entropy as $S=\ln K_g^{3/2}+$ constant.} 
  has been observed in the ICM of cluster cores \citep{Cavagnolo2009,10,Eckert2013}. Moreover, observations of the cool regions of  
galaxy clusters show very little evidence of gas cooling and  negative temperature gradient outside cores, which is more pronounced in the 
cool core (CC) clusters (clusters which have short cooling times, high gas densities and low temperatures in the central regions) \citep{Vikhlinin2005,Leccardi2008}.  Such issues can be addressed both through observations and simulations and these point towards cosmological scenarios where a number of non-gravitational astrophysical processes like star formation, 
feedback from supernova and Active Galactic Nuclei (AGN) take place. It has been 
found that radio emission from AGN feedback strongly correlates with the  
non-gravitational energy at cluster cores 
\citep{Chaudhuri2012,Chaudhuri2013,Pike2014,Planelles2014}. The presence of the 
cavities  in  bright galaxies \citep[e.g][]{Pandge2012} and in the ICM near 
the cluster centers \citep[e.g][]{Hlavacek2012,Hlavacek2015} is believed to 
be direct evidence of the conversion of the mechanical energy associated with 
the AGN jets into the thermal energy of the gas and possible small fraction of 
the non-thermal (relativistic) component.

Studies at radio wavelengths have also revealed diffuse radio sources (giant 
halos and relics), which cannot be generally associated with any 
central radio galaxy (AGN). Radio halos have extended structures ($\geq$ 1 Mpc) 
located at cluster centers while radio relics are elongated structures with 
an arc-like morphology mostly located at cluster periphery and are 
highly polarized. Such diffuse radio halos (giant halos, relics) can  be 
produced by the  re-acceleration of particles in fossil radio plasma,  or 
directly from thermal electrons of the ICM through shocks and/or 
magnetohydrodynamic (MHD) turbulence during cluster mergers - leptonic model 
\citep{Brunetti2001,Gitti2002,Cassano2005}.  They can be also produced from 
``secondary electrons'' as a result from inelastic collisions between cosmic ray 
(CR) protons and the thermal nuclei of the ICM  during cluster merger - 
hadronic model \citep{Dennison1980,Blasi1999}. 
 Mini-halos are also diffuse, faint radio sources mostly found 
at the cluster center and are normally of size $\sim100-500$ kpc comparable to 
that of the cluster cool core, with steep radio spectra.
Apart from being much smaller than giant radio halos which are typically found in  
disturbed/non cool core (NCC) clusters, radio mini-halos 
are quantitatively similar \citep{Brunetti2001}.  The importance of GMRT Radio Halo Survey 
and its extension stands out, which has provided us the statistical basis on the 
occurrence of giant radio halos in the critical
redshift range $0.2-0.4$ \citep{Cassano2008,Venturi2008,Kale2013,Kale2015a}. 
These results from the 
GMRT survey have provided observational support to the re-acceleration 
scenario for the formation of radio halos.

It has become clear that the studies from the large-area surveys like GMRT, VLA, LOFAR, SKA \citep{Condon1998,Haarlem2013,Blake2004} will provide us useful insights of the large-scale structure formation and in our 
understanding of cosmological model. The future SKA telescope programme, in  which India will also play a major part, is expected to detect thousands of galaxy clusters up to redshift $\sim0.6$ and several hundreds at redshift $z>1.5$ using radio sources associated with the AGN as well as ICM \citep{Gitti2014}. In this article, we  mainly focus on the radio emission from AGN and its effect on the surrounding ICM in galaxy clusters keeping in mind the physics that we can do with SKA.  The cluster detections via radio sources using SKA, in combination with  datasets in X-ray and cosmic microwave background radiation (CMB) \citep{Smoot1992} will provide us useful insights of the overall properties of clusters such as masses, luminosities
or temperatures, their deviations from the  self-similar scaling relations and in understand the regulation of cool cores \citep{Gitti2014,Combes2015}. Another major breakthrough with SKA will be in the investigation of radio-mode AGN feedback at high redshift which will help us in characterization of the radio lobes produced by the relativistic electrons which are responsible for carving the cavities in the ICM clearly observed in X-ray local clusters \citep{Hlavacek2012,Hlavacek2015}. The SKA will also be able to follow-up galaxy clusters using the Sunyaev-Zel'dovich (SZ)  effect \citep{Sunyaev1972,Sunyaev1980} signature from which one can in principle study the effect of the AGN feedback on the SZ power spectrum and 
thus the evolution of the large-scale structures by correlating the radio luminosity with 
thermodynamic properties of the ICM \citep{Chaudhuri2013,Grainge2015}.

\section{AGN feedback in galaxy clusters}
In order to obtain robust estimates of cosmological parameters from galaxy cluster surveys, one requires not only precise knowledge about the evolution of galaxy clusters with redshift but also precise determination of thermodynamical properties of the ICM. 
In the simplest cases, where one considers pure gravitational collapse, cluster scaling relations are expected to follow simple self-similarity \citep{Kaiser1986}. In this regard, X-ray scaling relations have been widely used to test the strength of correlations between cluster properties and to probe the extent of self-similarity of clusters \citep{Edge1991,Pike2014,Morandi2007,Comis2011,Ettori2013}.
For example, the luminosity-temperature ($L_x-T$) relation for self-similar models predict a
shallower slope ($L_x\propto T^{~2}$) than observed ($L_x\propto T^{~3}$) \citep{Edge1991,Andreon2011}. Such observations show that there is a break in the self-similarity in galaxy clusters, with a steeper slope, especially, for low mass clusters. Similarly, SZ effect scaling relations have also been largely studied both analytically and by numerical simulations, and these studies too show 
discrepancies between observations and prediction from a pure gravitational model \citep{Bonamente2008,Battaglia2012,Holderb2001,Zhang2003}.

Considering straightforward case, one may argue that radiative cooling may have a role for the steepness of the $L_x-T$ relation by eliminating gas more efficiently particularly in low-mass systems \citep{Lewis2000,Yoshida2002,Wu2002}
but this process produces a drastic overprediction of the amount of gas cooler than 1-2 keV which is clearly not consistent with the present optical and X-ray observations \citep{Balogh2001}. Another observational mystery  is the ``cooling-flow'' problem in the galaxy cluster. Since it is found that the cooling time in the  galaxy clusters cores is smaller than the age of the cluster, a central inflow of cool gas called cooling flow is expected to occur \citep{Cowie1977,Fabian1977}. Under these conditions, the gas slowly loses pressure support and falls towards
the central galaxy. However, it has been found that the estimated rate of gas cooling and accretion onto the central galaxy is rather high in many cases \citep{Fabian1984} suggesting that it is simply not the case \citep{McNamara2000,Peterson2003}. Secondly,  low star formation rates of central galaxies \citep{Johnstone1987,Connell1989,Allen1995} provide additional proof for the absence of significant amount of cooling in cluster cores. 
In addition, recent observations with radio and X-ray telescopes have also revealed complexity of the ICM physics, such as cold fronts, radio ghosts, cluster turbulence and nearly uniform high metallicity.

All these findings suggest that there must be a physical process that offsets the radiative cooling in the cluster cores, thus preventing the gas from falling out of the ICM in a cooling flow.
Such studies has revealed the importance of complex non-gravitational processes, such as injection of energy feedback from AGN, radiative cooling, supernovae, and star formation, influencing the thermal structure of ICM, particularly in low mass (temperature) clusters
\citep{10,Eckert2013,Chaudhuri2012,Chaudhuri2013,Voit2002,Voit2005}. The first direct evidence for non-gravitational entropy in galaxy clusters and galaxy groups was given by \cite{David1996} using ROSAT PSPC observations.
\cite{Ponman1999} found flatter entropy profile in galaxy cores and \cite{Helsdon2000} results indicated much steeper slope ($\sim4.9$) for luminosity-temperature relation in galaxy groups than observed in galaxy clusters. 
 Motivated by these findings several groups have reported similar conclusions using both numerical and semi analytical models  with an entropy floor of the order of $200-400$ keV cm$^2$ \citep{Eckert2013,Bialek2001,Borgani2005}.
By investigating the effect of excess energy on the density/temperature profiles of ICM gas using analytic/semi-analytic models, the excess energy per particle 
has been estimated to be around $0.5-3$ keV \citep{Wu2000,Chaudhuri2012,Chaudhuri2013}. However, the main unsolved issue in these models remains the origin and nature of the physical sources that cause the extra heating of the ICM. 

Although, Supernovae feedback is essential to explain the enrichment of the ICM to the observed metallicity level and heavy-element abundances but it provides insufficient estimates of energy per particle ($< 1$ keV) as compared to recent observations. Moreover, it is also inefficient to quench the cooling in massive galaxies  \citep{Springel2005}. An older idea to explain excess entropy/energy  is that of preheating. This proposes that the cluster forms from an already preheated and enriched gas due to feedback processes (like galactic winds) heating up the surrounding gas at high redshifts. The preheating model was first proposed by  \cite{Kaiser1991} and has since been developed/improved by many \citep{Borgani2001,Evrard1991,Tozzi2001,Babul2002,Finoguenov2003}. The simplest preheating scenarios require $\sim1$ keV energy per particle or constant entropy floor of  $> 300$ keV cm$^2$  along with the  radiative cooling  to explain break in the self-similarity scaling relations \citep{Babul2002,McCarthy2002,Iqbal2017a, Iqbal2017b}. 
Although, early preheating models could describe the scaling relations in clusters, they suffered from a few drawbacks with regard to details. 
For example, such models predict isentropic cores particularly in the low mass clusters \citep{Ponman2003} and the excess of entropy in the  outskirts of the clusters \citep{Voit2003} which are not consistent with observations.

There is a growing evidence that AGN feedback mechanism provides a major source of heating for the ICM gas, thereby reducing the number of cooling flow clusters \citep{Nath2002,Roychowdhury2005,Chaudhuri2012,Chaudhuri2013,Pike2014,Planelles2014,Guo2008,Gaspari2011,Daalen2011}.
The AGN-jet simulations have reached a stage where they can suppress the cooling flow for a cosmological timescale and produce results matching with observations \citep{Gaspari2012,Li2015,Prasad2015}. Understanding the physics of the hot gas and its connection with the relativistic plasma ejected by the AGN is key for understanding the growth and evolution of galaxies and their central black holes, the history of star formation, and the formation of large-scale structures.  The effect of AGN feedback can occur at different scales ranging from galaxy formation to cluster cores. In case of galaxy formation, it reduces the galaxy luminosity function by suppressing the over-production of massive elliptical galaxies as predicted by dark matter only simulations, while in case of cool core systems it solves the cooling flow problem. Additionally, it also gives helpful insights about the observed relation between the black hole mass and the bulge velocity dispersion \citep{Gebhardt2000}. Current observations suggest advection dominated accretion flow (ADAF) systems as the primary power source behind local AGNs \citep{Narayan2008}. Studies from high resolution Chandra X-ray telescope have found that  AGN can inject up to about $10^{58}-10^{62}$ erg per outburst into the ICM  which is not only sufficient to quench the cooling of the ICM especially in CC clusters and increase its energy but also suppresses star formation and the growth of luminous galaxies \citep{Birzan2004,McNamara2007}.

However, many questions regarding the  the physics of AGN feedback and how it interacts with the surrounding ICM, remain unanswered.
Several different mechanisms have been put forward and have been investigated numerically and through observations. These include radiative heating by quasars called ``radiative feedback'' \citep{Ciotti1997,William2007} or bipolar mechanical outflows/jets 
called ``kinetic feedback'' \citep{Gaspari2011,Zanni2005} While quasars might have been an important source of heating at high redshift, with the
peak distribution at $z \sim 2$ \citep{Nesvadba2008,Dunn2010}, however, in low-redshift systems it seems likely that massive black holes mostly accrete mass and return energy in a radiatively inefficient way \citep{Fabian1995}.
The detection of radio-filled X-ray cavities/bubbles, lobes, ellipsoidal weak shocks and iron enhancements along the radio jet trajectory strongly suggests that low redshift AGNs introduce energy directionally  and in mechanical form through bipolar massive jets. Recent ALMA observations \citep[e.g.,][]{Tremblay2016} and theoretical models (e.g., \citealt{Hobbs2011,Gaspari2013,Prasad2016}) suggest that accretion inward of $\sim 1$ kpc occurs in the form of colliding cold clouds and a turbulent cold (but thick) disk, but  many missing gaps in theory/simulations still need to be understood.

A recent promising development in understanding the state of the ICM is based on the concept of local thermal instability in core which is  in rough  global thermal balance. In isobaric conditions of the ICM, slightly denser blobs of the ICM are expected to cool faster than  their surroundings. However, this local thermal instability results in cold gas only if the ratio of the ICM cooling time and the gravitational free-fall time is less than a critical value close to 10 \citep{Sharma2012a,Meece2015,Choudhury2016}. This criterion quantitatively explains the observed entropy threshold ($\lesssim 30$ keV cm$^2$) for the presence of H$\alpha$ luminosity and radio bubbles \citep{Cavagnolo2008}. Local thermal instability picture ties together with the cold mode feedback as opposed to hot-mode Bondi accretion  \citep{Pizzolato2005,Pizzolato2010} which explains several observations \citep{Voit2015}. Thermal instability and condensation model also has important implications for the $L_x-T$ relation and the missing baryons problem \citep{Sharma2012b}.

The synchrotron radiation emitted by the relativistic electrons in radio bubbles fades and becomes difficult to detect after about $10^8$ years. 
Moreover, the  X-ray surface brightness depressions are only visible near the center of the cluster 
where the contrast is large. Thus, it is unclear how far AGN driven cavities rise in the cluster, their interaction with the ICM, 
and how they evolve at late stages.
In order to gain more physical picture of the nature of interactions between the AGN feedback and ICM 
and the extent of balance between mechanical power fed by the AGN versus radiative loss of ICM, it is important to  explore X-ray bright systems with apparent signatures of such interactions. Upcoming X-ray and radio observations could help us to narrow down search for physical feedback scenarios \citep{Gitti2012}. Studies of X-ray deficient cavities allow us to derive the relationship between the mechanical energy injected and radio emission of AGN jets and lobes. Such a relationship is of great interest because it can helps to understand the physics of AGN jets \citep{O'Sullivan2011}.

The SKA1-mid and SKA2-mid  observations would allow us the imaging of jets up to $z\approx 10$ which will help us to accurately estimate physical conditions in the jets. This will permit us to precisely determine the variations in the pressure and density along the jets  and their dependence on jet power and star formation \citep{Laing2002}. Considering a baseline of about 100 kms SKA1-mid Band 4 (centered on 4 GHz and bandwidths 2.4 GHz) and Band 5 (centered on 9.2 GHz and 2 $\times$2.5 GHz) will help us to resolve  images of jets and lobes for both young and evolved radio galaxies at 0.4-0.07 arcsec resolution. One can in principal, therefore, do the detailed study of the spectral evolution of the fading radio lobes both in synchrotron and Compton regime in the varying magnetic field which drives the expansion \citep{Goldshmidt1994}. Owing to high  sensitivity and frequency range, SKA will also be able to trace the evolution of inflated bubbles to much larger distances in the cluster. This will offer us unique opportunity to study the physical properties of relativistic gas inside the bubbles, such as the buoyancy, equation of state, magnetic field and hydrodynamical instabilities in bubble and their interaction with the ICM \citep{Scannapieco2008a,Scannapieco2015}.  

The SKA will also contribute in explaining much debatable mechanism responsible for the radio emission in radio-quite AGNs where the emission is mostly dominated by the thermal emission related to accretion disk unlike radio-loud AGNs where the spectrum is dominated by non-thermal spectrum from relativistic jets. The detailed knowledge of physical processes occurring in the radio-quiet AGNs will not only be crucial for understanding their differences  with the  radio-loud  AGN population  but also in investigating a possible link between the AGN and the star formation. Using SKA-mid (2-5 Band), the angular resolution of 0.4-0.07 arcsec will be enough to separate the AGN emission of the radio-quiet AGNs from the star forming regions up to the $z\approx2$ \citep{Padovani2011,Orienti20014}. The SKA multi frequency spectrum  along with the polarimetric information can also help us in distinguishing the spectrum from different radio regions.  For example,  polarization information could help us to disentangle AGN jet emission from  star-forming regions where the spectrum is expected to be highly polarized  \citep{Agudo2014}.

\section{Probing AGN feedback in galaxy clusters with SKA}
\subsection{Estimates of feedback energy and correlation with radio luminosity}
The thermodynamic history of the ICM, and hence any energetics, is fully encoded in the entropy of the ICM.  Using both numerical and semi analytical models, an entropy floor of the order of $300-400$ keV cm$^2$ \citep{Ponman1999,Tozzi2001,Finoguenov2003,Eckert2013} has been found in the ICM of  cluster inner regions which typically translates into excess feedback energy per particle between $0.5-3$ keV. For example, \cite{Lloyd-Davies2000} have shown from observations that excess energy per particle is $0.44\pm0.3$ keV in groups of clusters while \cite{Borgani2001} have shown using  numerical simulations that one needs  excess energy of order $\sim1$ keV per particle to reproduces the observations.

Non-radiative simulations, which encodes only gravitational/shock heating, predict entropy profiles of the form $K(r) \propto r^{1.1}$ \citep{Voit2005}.
By comparing the observed entropy profiles ($K_{g,obs}$) with theoretically expected entropy profiles ($K_{g,th}$), based  on non-radiative gravitational/shock heating simulations, one can determine the nature and degree of feedback. Considering a transformation from the baseline configuration (non-radiative model) to new (observed) configuration 
i.e $\Delta K_{\textrm{ICM}}=K_{g,obs}-K_{g,th}$, the additional energy per particle in ICM corresponding to the transformation is given by \citep{Chaudhuri2012,Chaudhuri2013},
\begin{eqnarray}
\Delta E_{\textrm{ICM}} &=& {kT_{obs} \over
(\gamma-1 )} {\Delta K_{\rm ICM} \over K_{g,obs}}  \qquad\qquad \quad \quad ({\rm isochoric})\nonumber\\
&=&{kT_{obs}  \over (1-{1 \over \gamma})} 
{  \beta ^{2/3} (\beta -1) \over (\beta^{5/3}-1)}
{\Delta K_{\textrm{ICM}} \over K_{g,obs}} \quad ({\rm isobaric}),
\label{eq:delq}
\end{eqnarray}
where $\beta=T_{obs}/T_{th}$, $T_{obs}$ and $T_{th}$ being the observed and theoretical (non-radiative model) temperatures respectively and $\gamma=5/3$ is the adiabatic index. For the extreme case, where $\beta=2$, the two above mentioned estimates of energy input per particle differ by only a factor of $1.14$.
The total feedback energy per particle in ICM is then obtained by adding the energy lost due to cooling  which had remained in ICM i.e,
\begin{equation}
\Delta E_{\textrm{feedback}}= \Delta E_{\rm ICM}+ \Delta L_{bol}\,t_{age},
\label{energy}
\end{equation}
where $\Delta L_{bol}$ is the bolometric luminosity emitted by the ICM in a given shell and $t_{age}\sim5$ Gyr is the age of 
the cluster.

\begin{figure*}
\centering
\begin{minipage}{7.05cm}
 \includegraphics[width = 7.05cm]{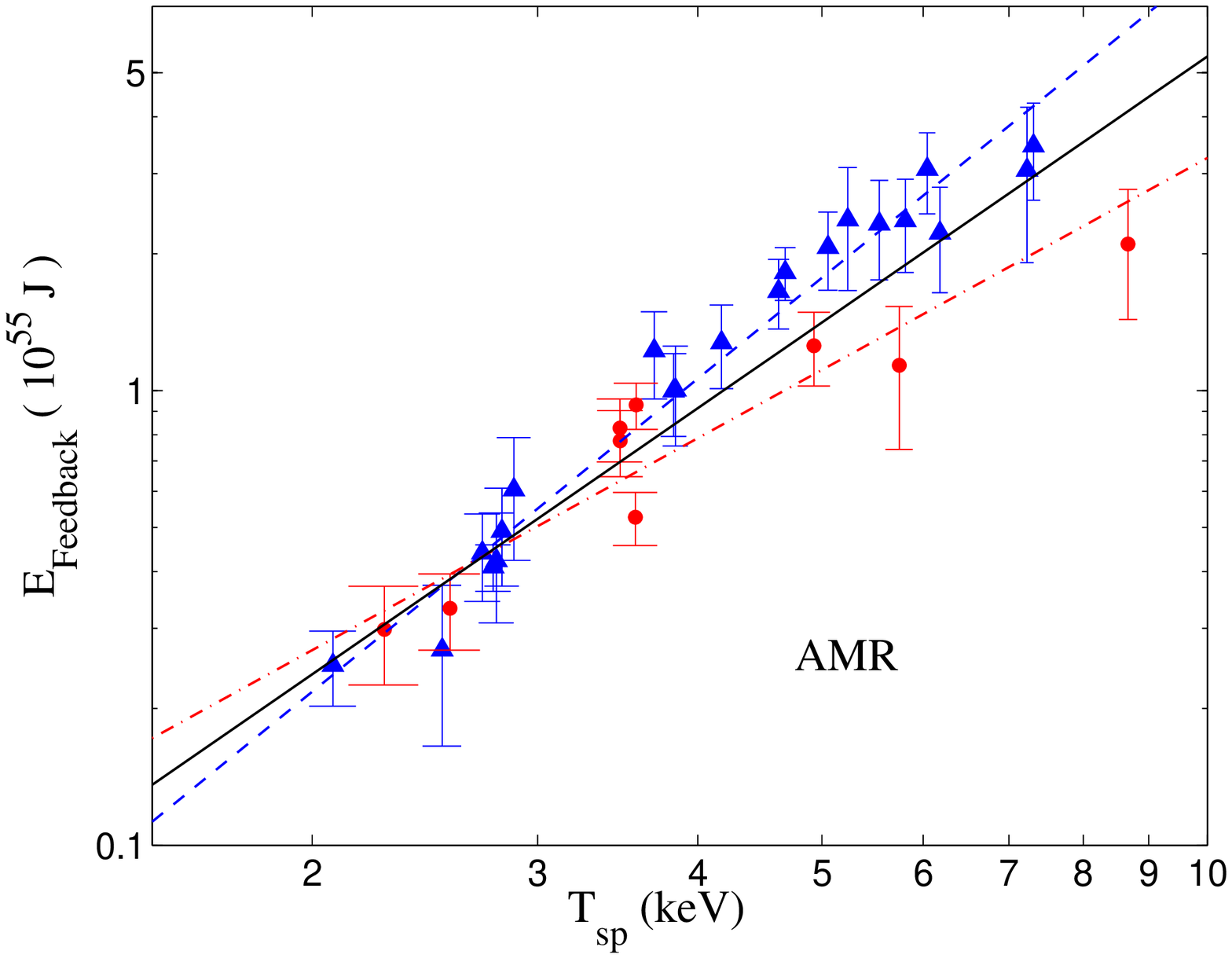}
\end{minipage}
\begin{minipage}{7.3cm}
 \includegraphics[width = 7.3cm]{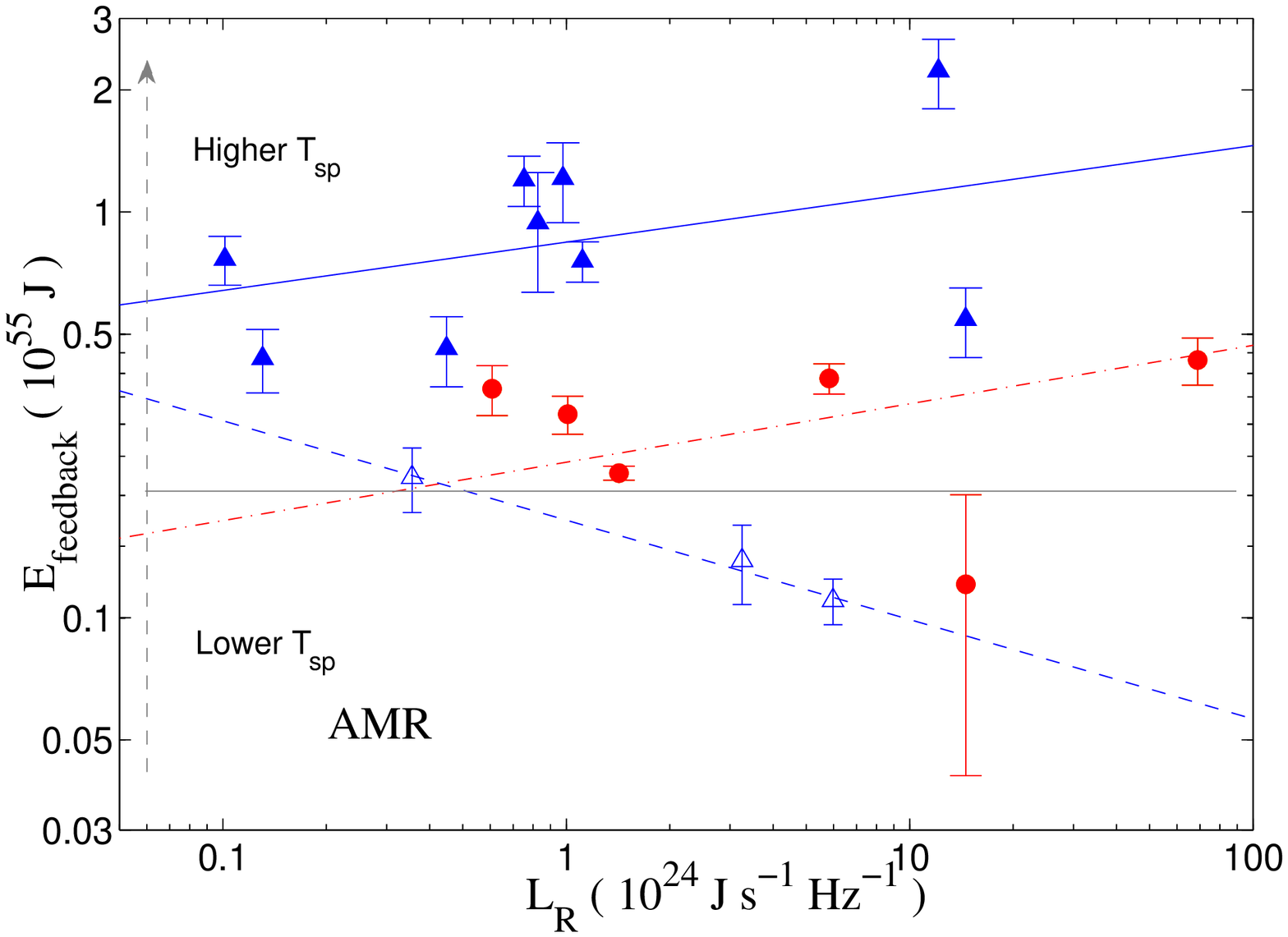}
\end{minipage}   
\caption{Left panel: Non-gravitational feedback energy $\Delta E_{\textrm{feedback}}$ verses cluster mean spectroscopic temperature $T_{sp}$ for AMR baseline entropy profile. The best-fit lines for the NCC clusters, the CC clusters, and the full sample are shown by the blue dashed line, the red dot-dashed
line and the black solid line, respectively. Right panel: feedback $\Delta E_{\textrm{feedback}}$ verses  radio luminosity  $L_{R}$ for AMR baseline entropy profile. The blue solid
and dashed lines are the best fits for the NCC clusters with  spectroscopic temperature $T_{sp} >$ 3 keV and $T_{sp} < 3$ keV respectively and the red dot-dashed line is the best fit to the CC clusters. Notice opposite correlation shown by low temperature clusters ($T_{sp} < 3$ keV). Figures adapted from \cite{Chaudhuri2013}.
}
\label{fig1}
\end{figure*}

Recently \cite{Chaudhuri2012,Chaudhuri2013} estimated the non-gravitational energy deposition profile up to\footnote{$\Delta$ is defined such that $r_{\Delta}$ is the radius out to which mean matter density is $\Delta \rho_c$, where $\rho_c=3H^2(z)/8\pi G$ 
being critical density of the universe at redshift $z$.} $r_{500}$
 by comparing the observed entropy profiles with a theoretical entropy profile without feedback \citep{Voit2005} for the  REXCESS sample of 31 clusters \citep{10}, observed with XMM-Newton. It is important to mention that unlike previous estimates they compare the feedback profiles at the same gas mass  ($m_g$) instead of same radii in order to take redistribution of gas on account of feedback processes \citep{Li2010,Nath2011}.
They found an excess mean energy per particle to be $\sim 1.6$ keV and  $\sim 2.7$ keV up to $r_{500}$ using theoretical entropy for AMR (adaptive mesh refinement) and SPH (smoothed particle hydrodynamics) simulations respectively showing strong correlation with AGN feedback. 
They found, as shown in left panel of fig.~(\ref{fig1}), that the total energy deposition corresponding to the entropy floor is proportional to the cluster temperature (and hence cluster mass) and scales with the mean spectroscopic temperature ($T_{sp}$) as $E_{\textrm{feedback}} \propto T_{sp}^{2.52\pm0.08}$ and $E_{\textrm{feedback}} \propto T_{sp}^{2.17\pm0.11}$ for the
SPH and AMR baseline profiles respectively.
They also observed the profiles of non-gravitational energy $\Delta E_{\textrm{ICM}}$ for CC and NCC clusters to be very different and it reached much lower values for CC clusters in the innermost regions as a
result of a greater amount of energy lost due to radiative cooling in CC clusters. However, adding energy lost due to cooling, the mean $E_{\textrm{\rm{feedback}}}$ profiles for CC and NCC appear to become more or less similar. 

By calculating the radio luminosity $L_R$ of the central radio sources within $0.3r_{500}$  from the NVSS catalog, \cite{Chaudhuri2013} found that this quantity is correlated with the bolometric luminosity $L_X$ for both CC and NCC clusters.  Moreover, their study also found that $L_R -\Delta E_{\textrm{feedback}}$ relation shows a strong trend for both the CC and NCC clusters with similar
power law slopes for the CC and NCC clusters (although with different normalizations) even though $L_R$ in the CC clusters was found to be  much higher than NCC for same temperature. While the high temperature CC and NCC clusters show a positive relation between feedback energy and $L_R$, 
the three low temperature NCC clusters  in the sample were found to have opposite trend (see right panel of fig.~(\ref{fig1})).
All these results indicate that AGN feedback from the central radio galaxies must provide a significant component of feedback energy in both CC and NCC clusters. Unlike, \cite{Chaudhuri2013} where they estimated the feedback energetics in the cluster inner regions, \cite{Iqbal2017a,Iqbal2017b}
studied feedback profiles in the cluster outskirts and showed that the influence of AGN feedback is of little importance approximately beyond $r_{500}$, with the observational estimated non-gravitational energy per particle being fully consistent with zero at the cluster outskirts. They also showed that preheating models which require entropy floor of $\approx300$ keV cm$^2$ and feedback energy per particle of $\approx1$ keV to explain break in self-similar scaling relation, is ruled out at more than 3$\sigma$.

A combination of the SKA arrays and their receivers at a wide range of frequencies and angular resolutions will be ideal for this kind of study. According to the results of \cite{Chaudhuri2013}, radio-loud AGNs with luminosity $10^{23}$ J s$^{-1}$ Hz$^{-1}$ at 1.4 GHz are important for low mass clusters, with X-ray luminosity $\le 10^{44}$ erg s$^{-1}$. These radio-loud AGNs are also the most abundant according to the radio galaxy luminosity function. The flux density of such a radio-loud AGN at $z=0.2$ is $\sim 1$ mJy, and at $z=0.5$, it is $\sim 0.1$ mJy. Therefore, SKA sensitivity will allow one to not only have a reliable estimate of the total radio power in low mass clusters at low redshift, but will also allow one to determine the redshift evolution of the role of radio-loud AGNs in clusters. In particular, this makes it  an indispensable tool for searching for old, dying radio sources, and for the construction of complete, flux density in the AGNs. The high sensitivity of the SKA along with its wide field of view  will help us in identification of objects of the same morphological type i.e  FR-I, FR-II or disturbed\footnote{Radio sources are often classified into two main classes, those having edge-darkened  morphologies (FR-I) and  those  having  edge-brightened   morphologies (FR-II) \citep{Fanaroff1974}.}. 
\subsection{Cool core clusters and radio mini-halos}
The X-rays emission from the hot ICM in the galaxy clusters represent a loss of energy. The time taken for the gas to radiate its enthalpy per unit volume ($H_v$) called the cooling time can be calculated as \citep{Peterson2006,Gitti2012,Fabian1984},
\begin{equation}
 t_{\textrm{cool}}\approx \frac{H_v}{\epsilon}=\frac{\gamma}{\gamma-1} \frac{k_BT}{\mu_g n_e \Lambda},
\end{equation}
where $\epsilon$ is the emissivity of the gas and $\Lambda$ is the cooling function. Since the cooling time in the cores of galaxy clusters is much smaller than the age of the clusters and a cooling rate is found to be high in cool core clusters, a central  cooling flow \citep{Sarazin1986,Bohringer2010} is expected to maintain the pressure required to support the weight of the overlying gas. The mass inflow rate, in a cooling flow model, can be estimated from the X-rays by using the luminosity $L_{\textrm{cool}}$ associated with the cooling and assuming steady-state isobaric condition \citep{Sarazin1986},
\begin{equation}
 L_{\textrm{cool}}=\frac{dE}{dt}=\frac{5}{2}\frac{\dot{M}kT}{\mu_gm_p},
\end{equation}
where $E$ is the total energy content in the ICM and $\dot{M}$ is the total mass deposition rate. However, from the  X-ray spectral observations it has been found that the gas cooling below  one third of its original temperature which is about 10 times less than expected from cooling flow models and is referred to as the cooling flow problem. 

Evidence gathered from the observations over several decades suggest the presence of central FR-I type radio galaxies in about 70\% of the cool core clusters \citep{Worrall1994,Komossa1999,Worrall2000}. With the advent of high-resolution X-ray observations 
using the present generation X-ray telescopes, it is well established that the central radio source produces a profound effect on the ICM. In particular, the X-ray images obtained at the superb spatial resolution using the Chandra telescope revealed that the hot gas in many cool core systems is not smoothly distributed, but instead exhibits highly disturbed structures including cavities or bubbles, shocks, ripples and sharp density discontinuities. Comparison of these features with the radio images at similar angular resolutions have revealed that these disturbances originated due to the AGN jets \citep{Blanton2001,Croston2003,Fabian2003}. The energy required to create a cavity is the sum of the internal energy ($E_{\textrm in}$) of the lobes and $pV$ work done by the jets to displace the ICM gas \citep{Sarazin1986,Gitti2012},
\begin {equation}
E_{\textrm{cav}} =E_{\textrm in}+ pV= \frac{\gamma}{\gamma -1}pV,
\end {equation}
where $p$ is the pressure measured in a shell (which is assumed to be in pressure equilibrium with the cavities) and $V$ is the volume of the cavity. Here, the value of $\gamma$  which is the mean adiabatic index of the gas inside the cavities is $4/3$ for relativistic case and $5/3$ for non-relativistic case. The total energy input is then $4pV$ for the relativistic case and $5/2pV$ in the non-relativistic case.  The quenching  of cooling flow is believed due to the heating through the dissipation of the cavity enthalpy and through shocks driven by the AGN outburst.

\begin{figure*}
\centering
\begin{minipage}{7.0 cm}
 \includegraphics[width = 7.0 cm]{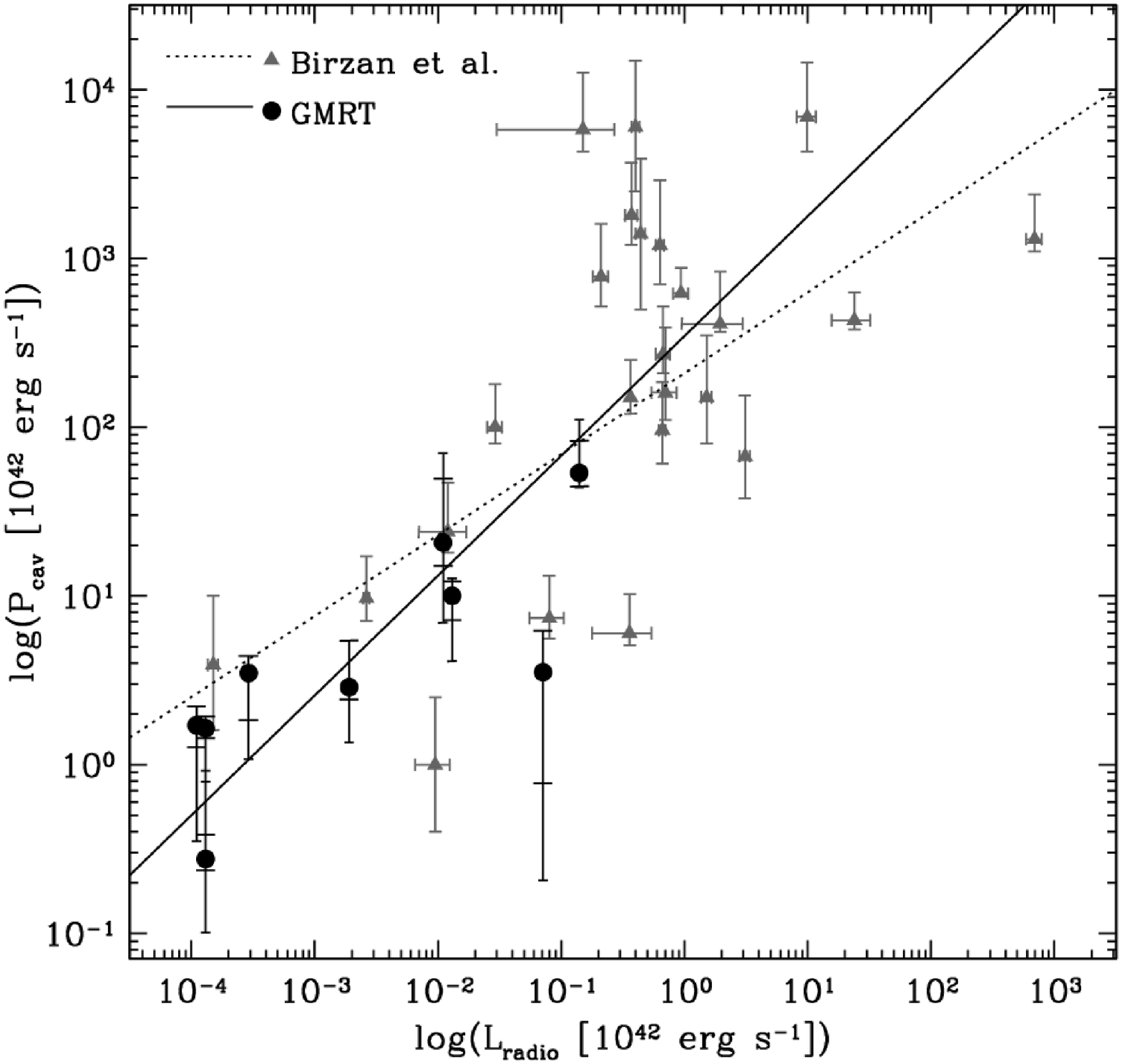}
\end{minipage}
\begin{minipage}{6.0cm}
 \includegraphics[width = 6.0cm]{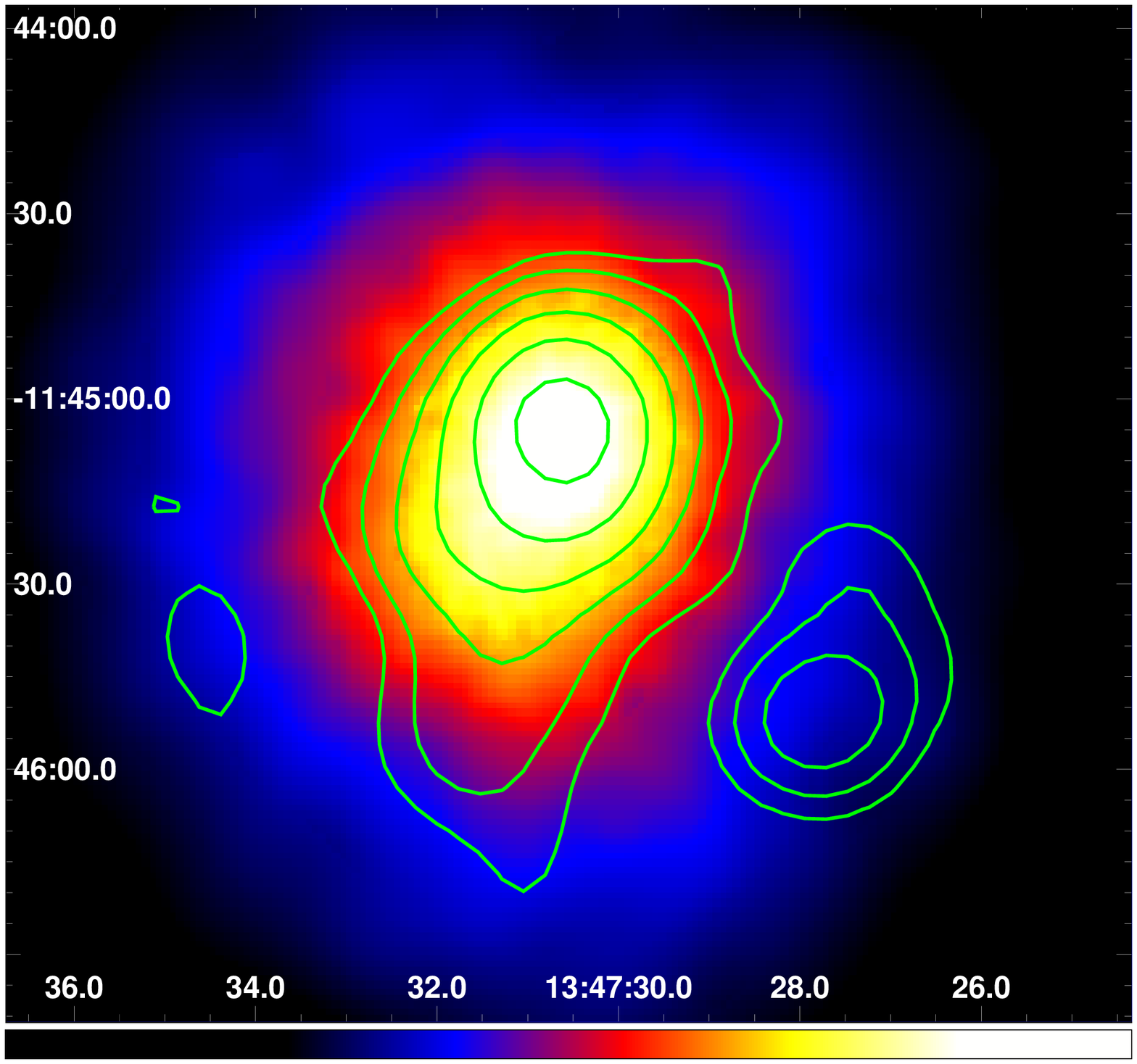}
\end{minipage}   
\caption{Left panel:  Cavity power vs integrated 10 MHz-10 GHz radio power. The solid and dashed line represent the best fit lines calculated by the \cite{O'Sullivan2011} and \cite{Birzan2008} respectively. Adapted from \cite{O'Sullivan2011}. Right Panel: 1.4 GHz contours of the radio mini-halo  in the galaxy cluster RXJ 1347.5-1145 ($z$ = 0.451), superimposed on the XMM-Newton image of the cluster. Adapted from \cite{Gitti2004}.}
\label{fig2}
\end{figure*}

In-depth analysis of individual systems along with the systematic studies of cavity samples \citep{Wise2007,Gitti2009,Rafferty2008,David2011} over a wide redshift range will enable us to investigate global properties of the X-ray cavities as well as shed light of the feedback mechanism in the cool core clusters.  Theoretically, it has been shown that these bubbles may rise buoyantly and raise some of the central cool gas \citep{Churazov2001}. Left panel in fig.~(\ref{fig2}) shows the  cavity power $P_{\textrm{cav}}$ verses integrated 10 MHz-10 GHz radio power $L_{\textrm{radio}}$.   Similar correlation of the radio power with surface brightness concentration, $c_{SB}$ (the cool core strength parameter which quantifies the excess emission in a cluster core) can be done using SKA and upcoming X-ray satellites like Wide Field
X-ray Telescope (WFXT) in order to study the evolution of cool cores which has been poorly understood till now because of the observational challenges of analyzing high redshift clusters \citep{Santos2010,Santos2012,Tozzi2015}. Large redshift sample will be crucial to establish cool core evolution and  their connection with the thermal properties and dynamics properties of ICM. Such studies would also help us in tightening the correlations involving the radio luminosity of the central AGN and the core entropy of the ICM.

Diffuse non-thermal emission has also been observed in a number of CC clusters, where the  radio-loud brightest cluster galaxy (BCG) is 
surrounded by a ``radio mini-halo''. Mini-halos  are believed to be generated from the ICM (instead of being connected with the radio bubbles) 
where the thermal plasma and the relativistic electron population are mixed \citep{Gitti2002,ZuHone2013}. The radio emission from the mini-halos can be due to relativistic electrons re-accelerated by MHD with the necessary energy supplied by the cooling flow process itself or due to the secondary electrons. Right panel of fig.~(\ref{fig2}) shows the radio emission overlayed onto the X-ray image  of the central region of RX J1347 \citep{Gitti2004,Gitti2007}. Study  of radio  mini-halos would provide valuable information on the micro-physical properties of the ICM, including the  MHD turbulence amplification of the magnetic field and the transport of CR in these environments. Only a handful of radio mini halos have been detected so far due to the limitations of current radio interferometers and difficulty in separating their diffuse, low surface brightness emission from the bright emission of the central radio source.  Due to these limitations it is difficult to understand the nature of the turbulence and  to discriminate between a leptonic and hadronic origin of radio mini-halos. However, owing to very good sensitivity to diffuse emission and good spatial resolution of SKA1 Band 2 (with a noise level of 2 $\mu$Jy), it is expected to detect thousands of mini-halos up to the $z\approx1$. This will allow us to reach a better understanding of their origin and physical  properties of these astrophysical sources and their correspondence with CC clusters.  For example, study of SKA observations in different bands along with Hard X-rays, gamma ray observations will help us in establishing the leptonic or hadronic model of mini-halos \citep{Gitti2014}. Such study will also enable us to establish a connection  between mini-halo origin and thermodynamical properties of ICM.
\subsection{Kinetic feedback of BCGs}
The BCGs or the first ranked galaxies in clusters are the most massive elliptical galaxies 
and show the highest probability to be radio loud. As said before, the BCGs impact the ICM through quasar mode and/or kinetic mode activity 
and are argued to be responsible for offsetting the cooling of cluster cores. However, the impact of the ICM on the BCG 
to regulate its radio duty cycle (which by definition is defined as the fraction of
time that a system possesses bubbles inflated by its central radio source) is still not well understood. 
The duty cycle can give us indirect insight of the AGN heating process and its inclusion in simulations could help us in understanding galaxy and cluster formation and evolution. Earlier work by \cite{Dunn2006} has estimated the  duty cycle of AGN heating in clusters to be 70\%. However,  \cite{Nulsen2009} have found much smaller value of duty cycle.  Although it has become increasingly clear, the role AGN feedback in regulating the cooling flows, it is still not well understood how the cooling time-scale relates to the AGN duty cycle. With wide spectral coverage of SKA (50 MHz-14 GHz),  it could be possible to determine duty cycle of AGN outbursts for  large sample which has been been found to be at least 60\% and could approach 100\% \citep{Birzan2012}.

\begin{figure}
\centering
 \includegraphics[width = 7 cm]{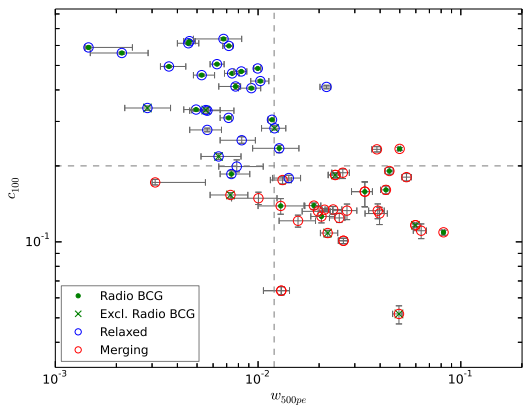} 
\caption{Distribution of the BCGs in the concentration parameter ($c_{100}$)- 
centroid shift ($w_{500}$) space. Relaxed clusters are shown as blue 
circles, and are all located in the upper
left quadrant; merging clusters are shown as red circles, and occupy the lower 
right portion. Filled circles are the radio loud
BCGs, the crosses show the BCGs removed from the analysis for reasons 
discussed in \citet{Kale2015b}. The threshold
values shown by dashed lines are to classify clusters as mergers or relaxed
according to \citet{Cassano2010}, i.e., $w_{500} > 0.012$ and $c_{100} < 0.20$.
Relaxed clusters show a higher occurrence of BCGs with radio emission. 
The figure is adapted from \citet{Kale2015b}.}
\label{bcgfig}
\end{figure}

It is also an open issue whether the large scale merging activity among   clusters can affect the radio loudness of the BCG. A recent study by 
\citet{Kale2015b} of the radio luminosity function of  a sample of 59 BCGs in the Extended GMRT Radio Halo Survey cluster sample ($0.2<z<0.4$, $L_{X[0.1-2.4 \mathrm{keV}]} > 5 \times 10^{44}$ erg s$^{-1}$)  \citep{Venturi2007,Venturi2008,Kale2013,Kale2015a}, has shown that the fraction  of radio loudness reaches about $90\%$ in BCGs for relaxed clusters as compared  to $30\%$ in BCGs for merging or non-relaxed clusters (fig.~(\ref{bcgfig})). The sample is not large enough to conclude if the shape of the radio luminosity function of BCGs in relaxed  and merging clusters is different. 

\begin{figure}
\centering
 \includegraphics[width = 7.0cm]{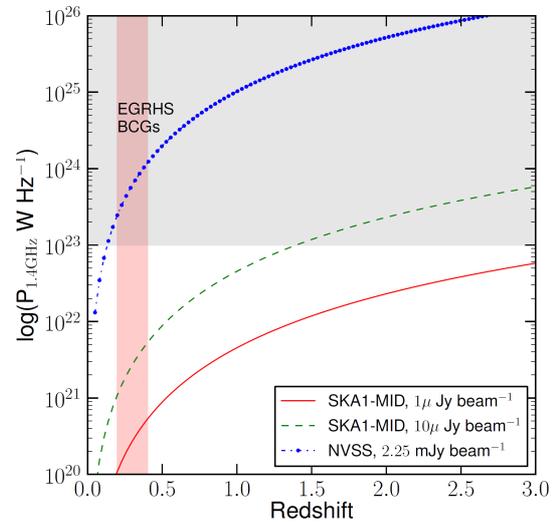}
\caption{The radio power at 1.4 GHz versus redshift for the detection thresholds of the NVSS (NRAO VLA Sky Survey) and 
SKA1-MID. The EGRHS (Extended GMRT Radio Halo Survey) sample of BCGs is limited to the redshift range of 0.2 - 0.4 shown by the 
vertical shaded band. The grey shaded region shows the typical range of radio powers of BCGs.}\label{bcgfig2}
\end{figure}

The all-sky surveys and pointed surveys of  galaxy clusters with the SKA-low and SKA-mid   will be able to provide large samples of radio BCGs in clusters to redshifts of 1 or even further.  It is clear from fig.~(\ref{bcgfig2}) that the SKA1-MID is expected to detect the BCGs in clusters up to redshift of 1.4 down to the power of $10^{23}$ W Hz$^{-1}$ with a fiducial source detection threshold of $10\mu$ Jy beam$^{-1}$. Single fields with deeper targeted observations can reach a factor of 10 deeper. The detection thresholds for the SKA1-MID are according to the proposed all sky ($3\pi$ steradians) and Deep Tier (2000 hrs) reference surveys \citep{Padovani2015}. These complete samples of radio BCGs, in combination with X-ray properties  can be used to understand if and how the large-scale properties of the ICM  affect the radio loudness of BCGs. 
 
 \subsection{Feedback in group scale environment}
 \begin{figure}
\centering
 \includegraphics[width = 7 cm]{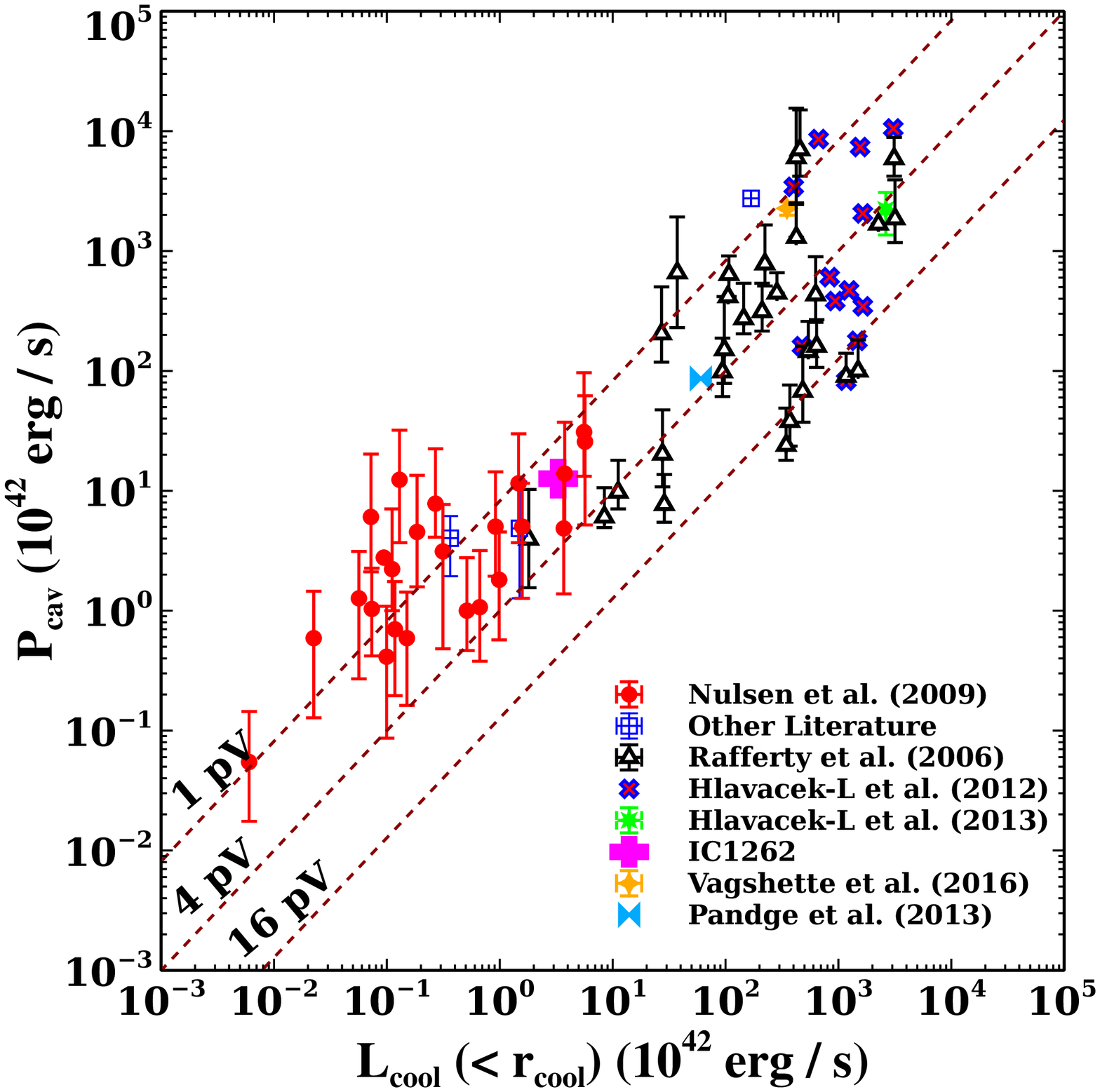} 
\caption{The correlation between the X-ray cavity power and X-ray cooling luminosity ($L_{\textrm{cool}}$) for the sample of groups. The diagonal lines represent samples where $P_{\textrm{cav}} = L_{\textrm{ICM}}$ assuming $pV$, $4pV$, $16pV$ as the total enthalpy of the cavities. Private communication Pandge, M. (2017). }
\label{fig3}
\end{figure}
Even though AGN feedback is well understood in the cooling flow clusters, however, due to the  characteristic differences between groups and clusters, knowledge of cavity physics in clusters may not be directly applicable for that in groups \citep{O'Sullivan2011}.  Systematic study of galaxy groups with AGN-IGM (intergalactic medium) interactions in the nearby region is important to understand the AGN feedback in relatively shallower systems because the relationship between AGNs and hot gas can significantly influence galaxy evolution in the group environment where much of the evolution of galaxies take place \citep{Eke20041,McNamara2012}.
Due to their shallower gravitational potentials, the AGN outbursts in such systems are also believed to produce a larger impact on the intra-groups medium in the form of  X-ray cavities and shocks. For example, \cite{Dong2010} showed the AGN feedback duty cycle depends on the size of the environment (i.e., galaxy clusters or galaxy groups). However,  \cite{Shin2016} have recently shown that X-ray cavities are similar among galaxy clusters, groups and individual galaxies, suggesting that the formation mechanism of X-ray cavities is independent of environment.

One  needs to cross check whether the AGN feedback can work efficiently in quenching the cooling flow in group scale environments. The X-ray cavities produced by the radio jets and IGM cooling luminosity within the cooling radius will give a clear picture about the mechanical work efficiency of radio jets in such shallower systems (see. fig.~\ref{fig3}).  
The groups selected from nearby regions can also be useful to probe regions closer to the central black hole with greater detail \citep{Forman2015}. With the help of X-ray and high precision radio continuum spectra from SKA, we can study the the merging history of the group, the characteristics of the cavities, gas stripping from galaxies and the roles of the central AGN  in heating the intra-group medium \citep{Rafferty2012}. Therefore, combination of X-ray data and observations with SKA will give a complete picture of AGN feedback in group scale environment.

\begin{figure*}
\centering
\begin{minipage}{8 cm}
 \includegraphics[width =8 cm]{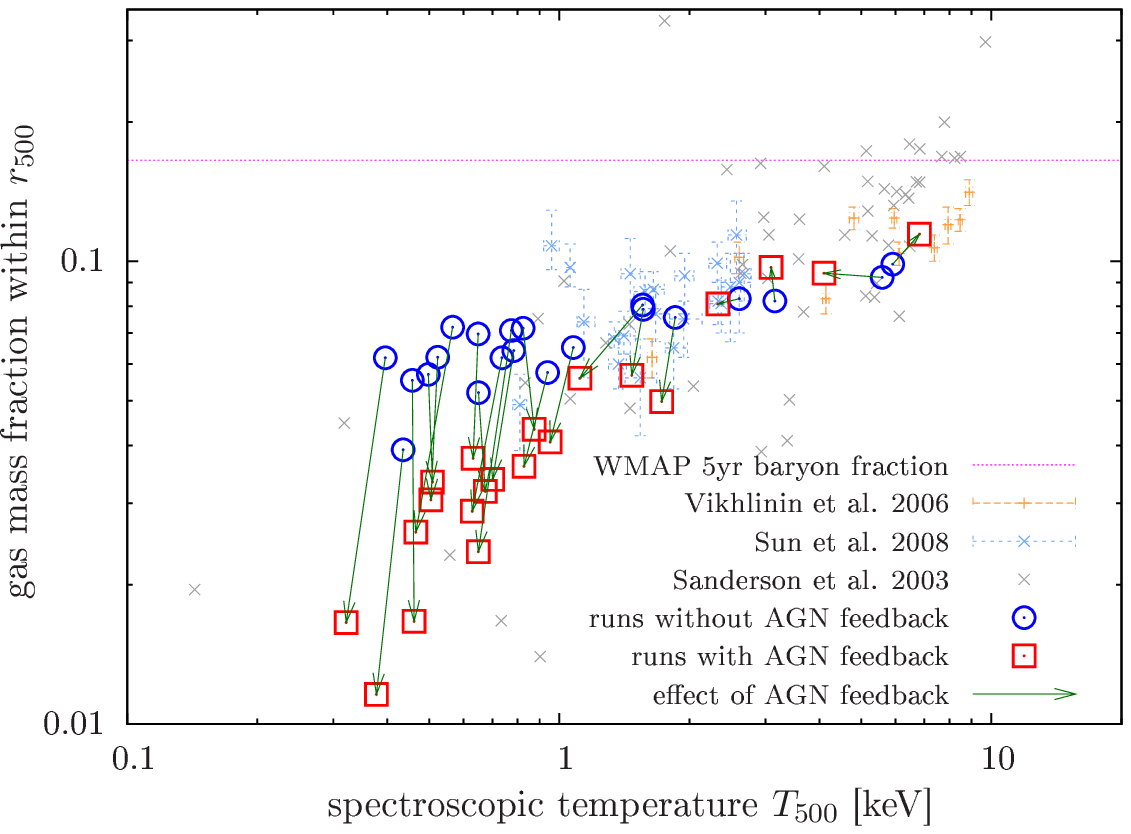}
\end{minipage}
\begin{minipage}{8cm}
 \includegraphics[width = 6.5cm]{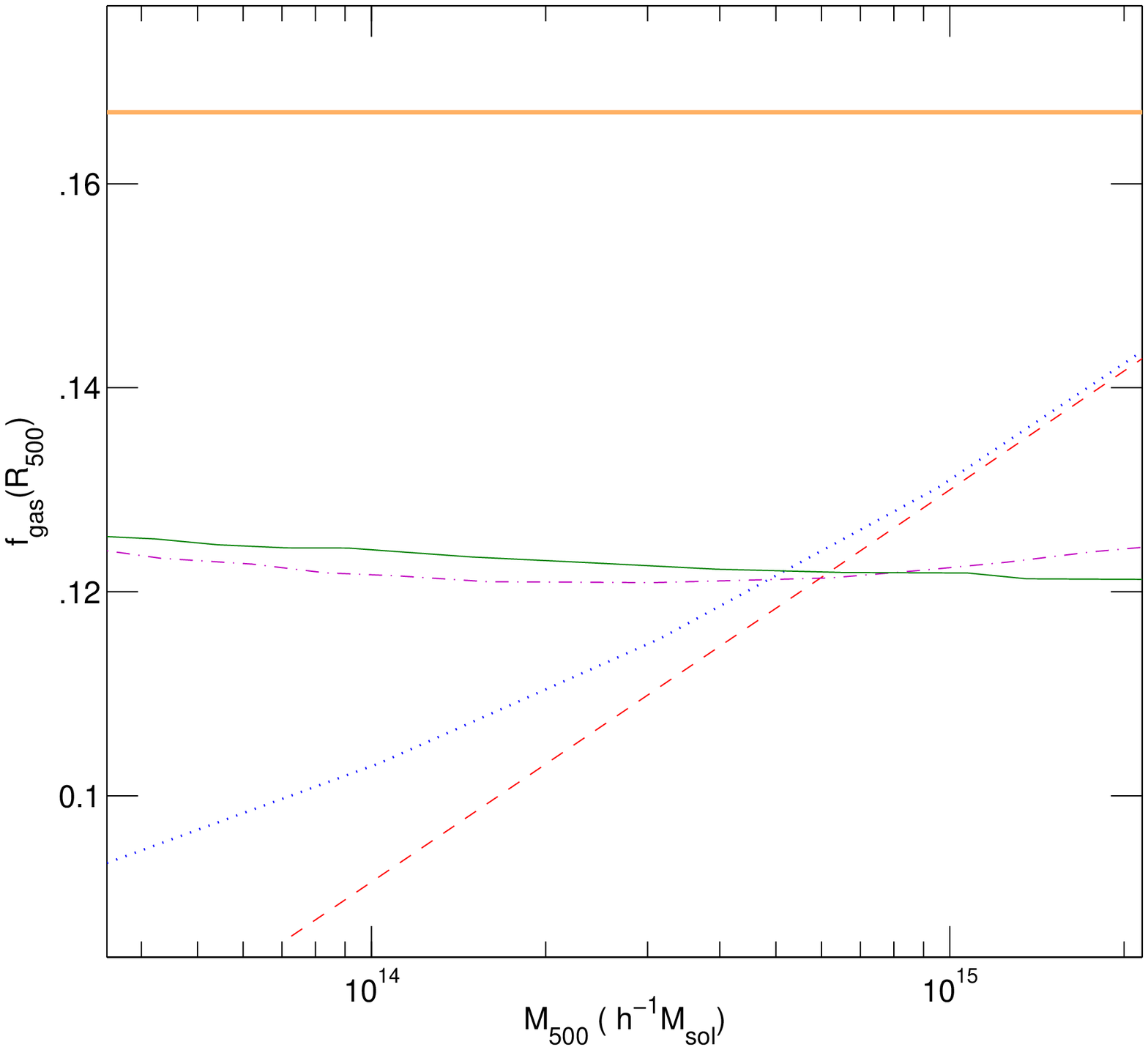}
\end{minipage}   
\caption{Left panel: Simulated gas mass fractions within $r_{500}$ of clusters (groups) without AGN feedback (circles) and with the feedback (squares) along with the observed X-ray estimates of different groups \citep{Sanderson2003,Sun2009}. The arrows illustrate the effect of the AGN heating for each halo. Figure adapted from \cite{Puchwein2008}.
  Right Panel: The $f_{gas}$ at $r_{500}$  plotted as a function of $M_{500}$.  The upper horizontal line is for $f_{gas}=\frac{\Omega_b}{\Omega_m}=0.167$; the solid and dot-dashed lines are for \cite{Komatsu2001} model for two different halo concentrations; the dotted line is for \cite{Chaudhuri2011} top down fiducial model and the dashed line is from X-Ray observations \citep{Vikhlinin2006}. Figure adapted from \cite{Chaudhuri2011}. }
\label{figf1}
\end{figure*}

\section{Implications for cluster cosmology - cluster physics synergies}
It has been pointed out that X-ray and SZ observables are sensitive to non-gravitational processes such as radiative cooling and AGN feedback and that the X-ray/SZ scaling relation have relatively large intrinsic scatter \citep{Edge1991,Pike2014,Morandi2007,Comis2011,Ettori2013}. Therefore, the utility of SZ surveys for determining cosmological parameters using the SZ power spectrum is limited by incomplete incorporation of energetic feedback from AGNs in modeling the gas (or gas pressure) profile in clusters. It becomes, therefore, imperative that for precision cosmology with X-Ray/SZ clusters, one needs not only the precise knowledge of cosmological distribution of clusters but also the thermodynamical properties of ICM.  Whereas, one could use self-calibration \citep{Majumdar2004} in large yield cluster surveys to treat scatter as a nuisance parameter to be marginalized over \citep{1}, a more direct approach is to reduce the scatter by introducing one extra parameter in the scaling relation which depends on the physics of the cluster center \citep{O'Hara2007}. This extra parameter can be better adjusted if one has additional knowledge of the AGN feedback at the  center.

There have been attempts from N-body-plus-hydro numerical simulations as well as analytic phenomenological models of the cluster gas to explore the effect of feedback in the gas distribution, for example, via the gas fraction and the resulting gas pressure profiles. \cite{Puchwein2008} performed high-resolution numerical simulations of a mass-selected sample of galaxy clusters (groups) to investigate the effect of AGN feedback on cluster (group) scales. They found that AGN feedback significantly lowers the gas mass fractions particularly in poor clusters and groups as feedback drives the gas from halo centers to their outskirts. It can be easily seen in left panel of fig.~(\ref{figf1}), where the simulation results of \cite{Puchwein2008} for gas mass fraction as a function of cluster X-ray temperature are shown for models with and without feedback. This significantly reduces the X-ray luminosities of poor clusters and groups (although temperature within $r_{500}$ stays roughly the same) which results in a steepening of the $L_X-T$ relation on the group scale.   Similar variation of gas fraction due to feedback has also been found by \cite{Chaudhuri2011} who built simple, top-down model for the gas density and temperature profiles for galaxy clusters. The gas was assumed to be in hydrostatic equilibrium along with a component of non-thermal pressure taken from simulations. The effect of the central AGN feedback was incorporated naturally using the slope and normalization of the concentration-mass relation, the gas polytropic index, and slope and
normalization of the mass-temperature relation.  This is shown in right panel of fig.~(\ref{figf1}) where the gas fraction deviated from the Universal gas fraction depending on the halo mass given by $f_{gas}(r_{500}$) = 0.1324+0.0284 $\log(M_{500}/10^{15} h^{-1} M_\odot$). The resulting gas profiles gave excellent agreement with both SZ and X-ray scaling relations. Moreover, joint study of radio emission and SZ signal has also become a promising tool to study the correspondence between the non-thermal and thermal component of the ICM \citep{Basu2012,Colafrancesco2014}. Joint analysis of GMRT observations (at 614 and 237 MHz) with high resolution MUSTANG instrument (90 GHz bolometric receiver on-board Green Bank Telescope (GBT)) results have already shown strong correlation between an excess in the radio surface brightness of the diffuse radio source at the center of the cluster  and a high pressure region detected in the SZ map of RX J1347  \citep{Ferrari2011}.

\begin{figure*}
\centering
\begin{minipage}{8 cm}
 \includegraphics[width =8 cm]{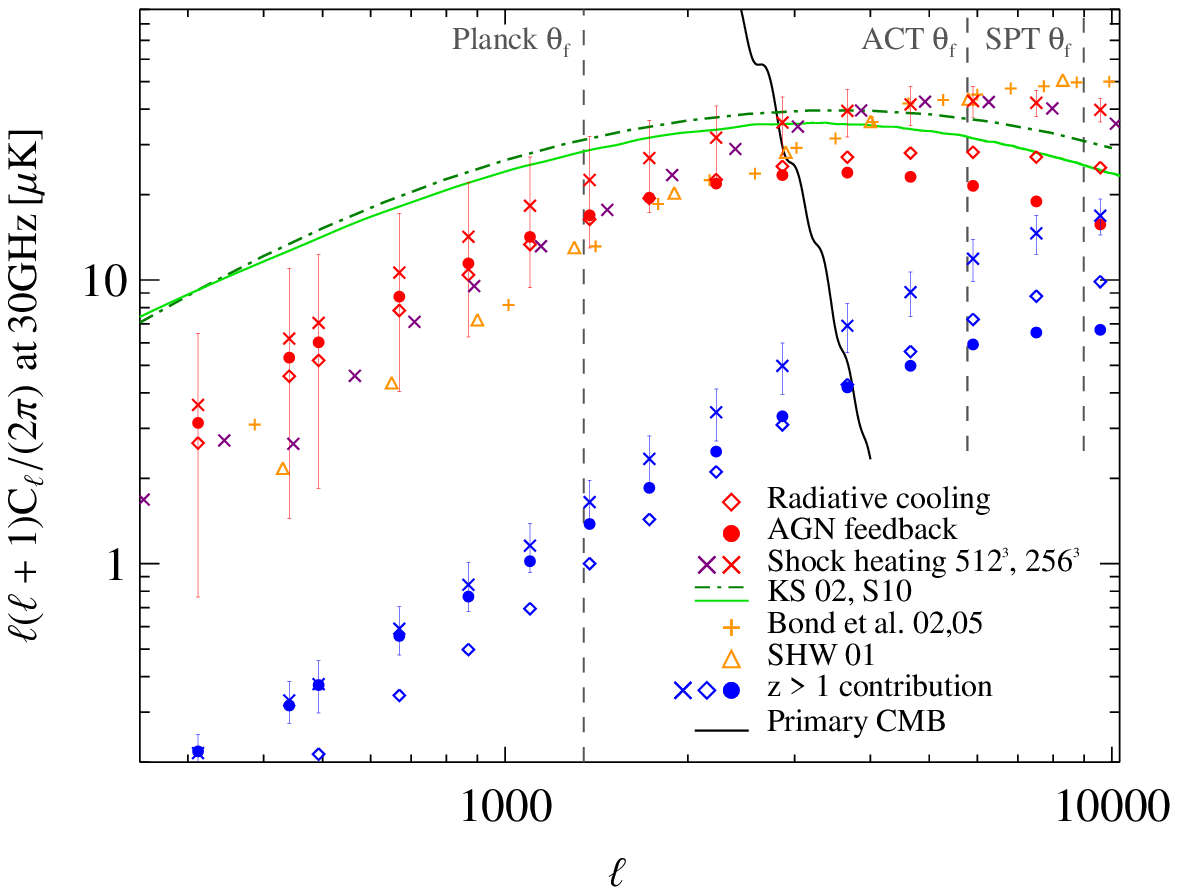}
\end{minipage}
\begin{minipage}{8cm}
 \includegraphics[width = 8cm]{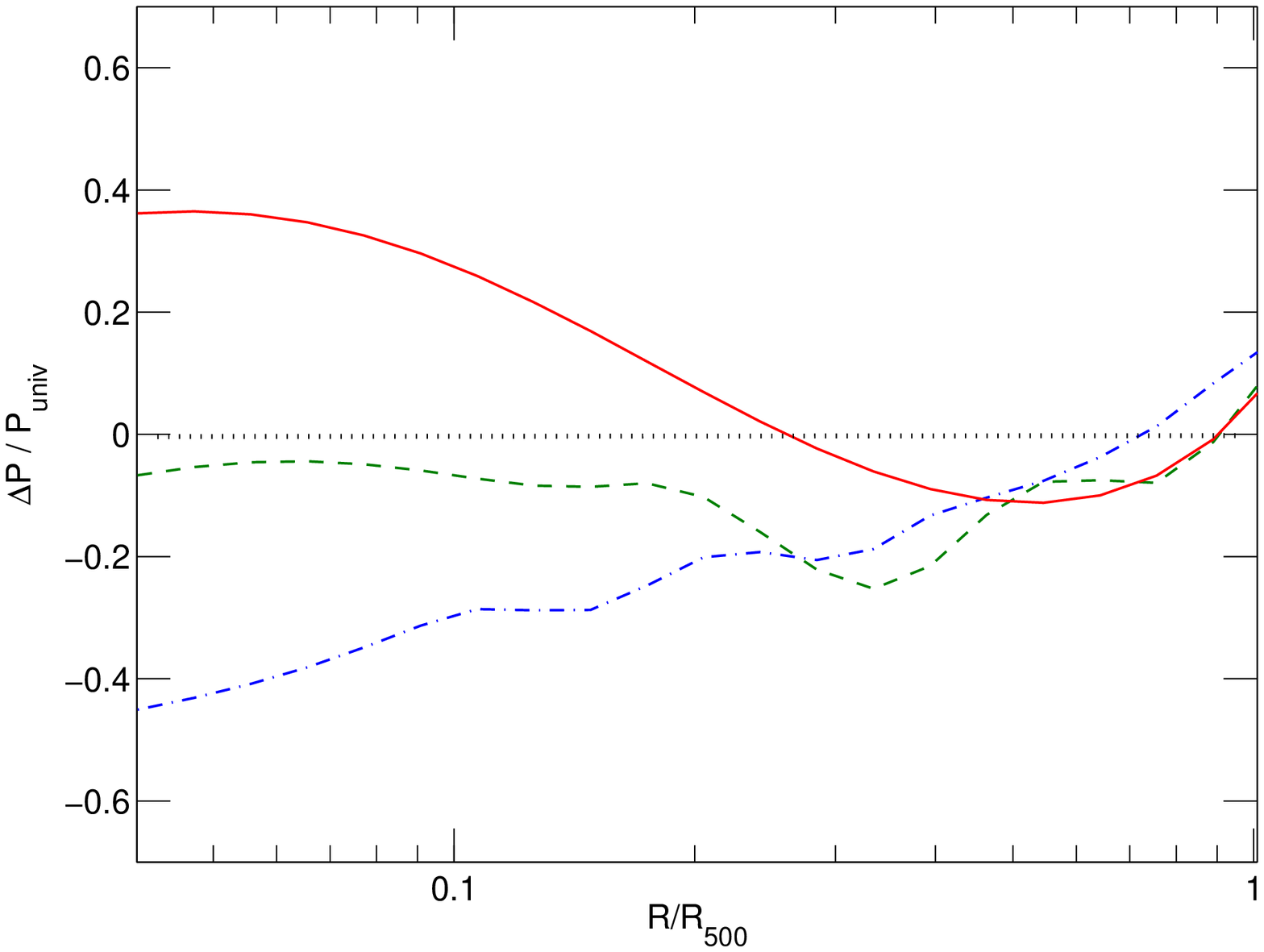}
\end{minipage}   
\caption{Left Panel:  SZ power spectrum at 30 GHz from \cite{Battaglia2010} simulation (red and purple symbols). AGN feedback model also include radiative cooling, star formation and SN feedback along with shock heating for the $256^3$ and $512^3$ simulations. Simulations by 
   \cite{Springel2001} (SHW01) and \cite{Bond2002,Bond2005} are shown with orange triangles and orange pluses respectively. Semi-analytical simulations by  \cite{Sehgal2010} (S10) and analytical calculations by \cite{Komatsu2002} (KS02) are shown with dark green and light green respectively. Figure adapted from \cite{Battaglia2010}.
   Right Panel:  Fractional pressure differences of phenomenological cluster model developed by \cite{Chaudhuri2011} (red solid line), from simulations by \cite{Battaglia2010} (green dashed line) and \cite{Sehgal2010} (blue dot-dashed line) w.r.t  universal profile found by \cite{11} for a cluster having M$_{500} = 2\times10^{14}h^{-1}M_\odot$. Figure adapted from \cite{Chaudhuri2011}.}
\label{figf}
\end{figure*}

Inclusion of AGN feedback in theoretical models also has implications for cosmology using the SZ power spectrum \citep{Majumdar2001, DiegoMajumdar2004, Aghanimetal2008, Scannapieco2008b, Battaglia2010, Soergel2017}. For example, \cite{Battaglia2010} found that the AGN feedback included simulations produces better match with universal pressure profile and cluster mass scaling relations of the REXCESS X-ray cluster sample \citep{11,10}. They showed that the CMB power at multipoles $\ell \approx 2000-10000$ which is probed by ACT and SPT is sensitive to the feedback and hence can indirectly constrain the physics of intra-cluster gas particularly for high redshift clusters. Similar to \cite{Puchwein2008} they also found AGN feedback not only directly influences the gas at inside the core but also pushes gas outside the core and impacts the SZ signal from the outer regions of the clusters. They concluded that the apparent tension between $\sigma_8$ from primary CMB and from  SZ power spectrum can be lessened with incorporation of AGN. Left panel of fig.~(\ref{figf}) shows the comparison of SZ power spectrum obtained by \cite{Battaglia2010}  with the other simulations and analytical/semi-analytic calculations \citep{Komatsu2002,Sehgal2010, Bond2002,Bond2005}. One of the crucial test of the AGN feedback simulations (like \citet{Battaglia2010}) and simple but quick phenomenological analytical model (like \citet{Chaudhuri2011}) is to reproduce gas pressure profiles (whose line integral gives the SZ distortion) from detailed X-ray observations inside $r_{500}$. The right panel of fig.~(\ref{figf}) shows a comparison of  pressure profiles obtained by \cite{Chaudhuri2012} with the universal pressure profile from observations. As evident from the figures,  on average, the theoretical models differ from observations approximately at the 20-30\% level up to $r_{500}$. Calibration of the AGN feedback energy using SKA observations would help in better modeling of the SZ power spectrum. 

Therefore, it is of utmost importance to understand the nature and the extent of the  non-gravitational feedback in galaxy clusters out to the virial radius so as to properly calibrate the scaling relations and theoretical models. Obtaining a much  large sample of simulated AGN-heated clusters with observations it will be possible to accurately calibrate observational biases in cluster surveys, thereby enabling various high-precision cosmological studies of the dark matter and dark energy content of the universe. With major advances in X-ray sensitivity, spectral resolution and  high spatial resolution, in combination with future radio observations especially from SKA, one may be able to make breakthroughs in these aspects of cluster physics. 

The SKA  is expected to investigate galaxy clusters using the SZ signal at $\approx 4-14$ GHz (SKA1-mid) which will be useful to constrain the masses of clusters  at high redshift ($z > 1$) due to its redshift independence. While the 
mass proxy measurements from the X-ray satellite  eROSITA  and  weak lensing measurements in the Euclid survey will be limited to
$z<1$ clusters, the follow-up targeted observations from  SKA1-mid  will provide mass estimates  through $Y-M$ scaling\footnote{$Y$ is called the integrated Computerization parameter,  defined as, $Y=2\times \pi D_A^{-2}\int_0^Ry(r)rdr$ where $D_A$ angular diameter distance and $y$ is the Compton $y$-parameter, $y=\sigma_{T}/\left(m_{\mathrm{e}}c^{2}\right)\int_{0}^{\infty}k_\textrm{B}T_{\mathrm{e}}(l)n_{\mathrm{e}}(l) dl$, $\sigma_{T}$ being the Thomson scattering cross section and integration is along the line of sight}. for $z>1$. It has been  shown through simulations  that a 1000-hour SKA1-mid programme can follow-up all the  high-redshift  clusters in SZ  which will be detected with eROSITA \citep{Grainge2015}. On the other hand one can also use  radio mini halos/radio halos as tracers of the galaxy clusters and can determine cluster selection function \citep{Cassano2014}. It was indeed shown by \cite{Gitti2014} that SKA1 and SKA2 have a  potential to detect up to 620 and 1900 mini-halos respectively at redshift $z<0.6$. Similarly,  \cite{Cassano2014} found that with SKA1-low and SKA1-sur it would be possible to detect up to 2600 and 750 radio halos, respectively out to $z\approx0.6$ while as maximum number of halos detectable by LOFAR telescope and proposed EMU survey (using Australian Square Kilometre Array Pathfinder (ASKAP) telescope) would be only 400 and 260 respectively. Studies of statistical properties of radio halos/mini-halos are important since they are found to be connected with the cluster dynamics. For example  \cite{Cassano2013} showed that radio halo power at 1.4 GHz scales with cluster X-ray (0.1-2.4  keV)  luminosity  and   integrated SZ signal computed  within $r_{500}$ as $P_{1.4}\approx L_{500}^{2.1\pm0.2}$ and  $P_{1.4}\approx Y_{500}^{2.05\pm0.28}$ respectively.

\section{Summary}
In this review, we have made in-depth analysis of AGN  feedback in large scale structures, in general, and galaxy clusters, in particular, and discussed the  potential impact of such feedback  in using clusters as  cosmological probes. A robust explanation for the mechanism behind AGN feedback in galaxies, groups and clusters is essential in understanding galaxies, 
central black holes, the history of star formation and the evolution of the large-scale structure. We discussed recent developments regarding the AGN  
outbursts and its possible contribution to excess  entropy in the hot atmospheres of groups and clusters, and its correlation with the feedback energy in ICM.  The much improved multi frequency measurements  and high spatial resolution in near future due to SKA will allow us to study the radio jets up to $z\approx 10$ and radio haloes/mini-halos up to $z\approx 1$  and its combination with up-coming data from cluster surveys from  Planck SZ and eROSITA will have the unique ability to address the most important outstanding questions in this field of research. The new observational capabilities of SKA will be pivotal in addressing the following issues

\begin{itemize}
\item SKA will help us in understanding the nature of the radio emission of the AGN and how it interacts with the surrounding medium narrowing  down the feedback scenarios. The SKA will give us a unique opportunity to study physical properties of relativistic gas inside the bubbles  and the evolution of jets/lobes. 
The SKA Band 4 and 5, for example,  will help us to resolve  images of jets and lobes for both young and evolved radio galaxies at 0.4-0.07 arcsec resolution up to $z\approx 10$ or so. Such studies will also help us to understand the possible mechanism responsible for the radio emission in radio-quite AGNs \citep{Agudo2014}. 

\item The proposed sensitivity of the SKA will help us in getting reliable estimates of the radio power even for the low mass clusters at high redshifts which has been found to correlate negatively with the total feedback energy  within cluster cores \citep{Chaudhuri2013}.  The radio detection down to luminosity $10^{23}$ J s$^{-1}$ Hz$^{-1}$ at 1.4 GHz will be important for low mass clusters having flux density of  $\sim 1$ mJy at $z=0.2$ and $\sim 0.1$ mJy at $z=0.5$ which are within the limits of SKA. Thus SKA will be a vital tool to study the correlation between feedback energy in ICM and radio power, as well as its redshift evolution.

\item In-depth analysis of the radio cavities and jets along  high spatial and spectral X-ray images over the wide redshift range  will be crucial in studying the evolution of the cool cores.  Moreover, SKA1 and SKA2 is expected to detect up to 620 and 1900 mini-halos respectively at redshift $z<0.6$ \citep{Gitti2014} which have been currently found in only a handful of CC clusters. The large sample of mini-halo will allow us to reach a better understanding of their origin and physical properties of these astrophysical sources and establish connection between mini-halos and CC clusters.

\item The all-sky and pointed SKA surveys will be able to provide large sample of BCGs up to $z\approx$ 1 or greater. Whereas as the EGRHS (Extended GMRT Radio Halo Survey) sample of BCGs is limited to the redshift range of 0.2 - 0.4, with SKA1-MID it will be possible to detect BCGs in clusters up to redshift of 1.4 down to $\sim 10^{23}$ W Hz$^{-1}$ with source detection threshold of $10\mu$ Jy beam$^{-1}$. This will allow us to study how the physics of the ICM can affect the radio loudness and duty cycle in BCGs and its evolution.

\item  SKA along with X-rays studies will also allow us to study the feedback in galaxy groups with far greater accuracy than present. Such study will be important to understand AGN-IGM  interactions and how the feedback can influence galaxy evolution and merging processes in the group scales.

\item SKA will able to measure both synchrotron and SZ radiation, thus becoming an important tool for studying  thermal and non-thermal physics of the ICM.   It is expected that 1000-hour SKA1-mid programme can follow-up all the high-redshift clusters in SZ which will be detected with eROSITA \citep{Grainge2015}. The radio halos/mini-halo properties are also found to be connected with the cluster dynamics.  Such studies will help us to remove the observational bias in the SZ scaling relations and power spectrum after proper incorporation of AGN physics.

\end{itemize}

Hydrodynamical simulations of radio-galaxy evolution and its  impact on the surrounding medium are becoming more and more sophisticated and are now capable of fully incorporating AGN feedback which can be used in understanding observational biases. Models of jet/lobe evolution and feedback implementations in cosmological  simulations can, therefore, be tested with direct observations (from SKA) and would provide major breakthroughs in  understanding the co-evolution of AGN, galaxies and  the
large-scale structure of the universe.


\end{document}